\documentclass[twocolumn,showpacs,preprintnumbers,amsmath,aps,amssymb,nofootinbib]{revtex4-1}
\usepackage{graphicx}% Include figure files
\usepackage{dcolumn}% Align table columns on decimal point
\usepackage{bm}% bold math
\usepackage{color}

\begin{document}

\title{Statistical modeling of the fluid dual to Boulware-Deser black hole}% Force line breaks with \\

\author{J. L. L\'opez${}^{a,b}$} \email{jl\_lopez@fisica.ugto.mx}
\author{Swastik Bhattacharya${}^{a}$} \email{swastik@iisertvm.ac.in}
\author{S. Shankaranarayanan${}^{a}$}%
\email{shanki@iisertvm.ac.in} \affiliation{
  ${}^{a}$School of physics, Indian Institute of Science Education and Research (IISER-TVM)  \\
  Thiruvanathapuram 695106, India \\ \\
  ${}^{b}$Departamento de F\'{\i}sica, Divisi\'on de ciencias e
  Ingenier\'ias Campus Le\'on,
  \\ Universidad de Guanajuato, A.P. E-143, C.P. 37150, Le\'on, Guanajuato, M\'exico. \\
}

\date{\today}% It is always \today, today,

\begin{abstract}
  In this work we study the statistical and thermodynamic properties
  of the horizon fluid corresponding to the Boulware-Deser (BD) black
  hole of Einstein-Gauss-Bonnet (EGB) gravity. Using mean field
  theory, we show explicitly that the BD fluid exhibits the
  coexistence of two phases; a BEC and a non-condensed phase
  corresponding to the Einstein term and the Gauss-Bonnet term in the
  gravity action, respectively. In the fluid description, the
  high-energy corrections associated to Gauss-Bonnet gravity are
  modeled as excitations of the fluid medium.  We provide statistical
  modeling of the excited part of the fluid and explicitly show that it is
  characterized by a generalized dispersion relation which in $D=6$
  dimensions corresponds to a non-relativistic fluid.  We also shed
  light on the ambiguity found in the literature regarding the
  expression of the entropy of the horizon fluid. We provide a general
  prescription to obtain the entropy and show that it is indeed
  given by Wald entropy.
\end{abstract}
% PACS, the Physics and Astronomy                             % Classification Scheme.
\maketitle

\section{Introduction}
  
Many interesting features of gravity have arisen since the formal
relation between the laws of thermodynamics and the laws of black hole
dynamics were found \cite{BCH,Bek,Hawking1}. These relations allow the
possibility of extracting information about the microscopic degrees of
freedom by providing statistical mechanical description of the
macroscopic properties of black hole horizons. [In this work, by degrees of 
freedom we mean the microscopic degrees of freedom corresponding to 
the black-hole entropy.] In other words, one
aims to arrive at the microscopic features of black holes from their
semi-classical properties as we yet lack information about the quantum
degrees of freedom of gravity. 
     
Fluid/Gravity correspondence is another approach that aims to
associate fluid degrees of freedom to the horizon and, eventually, to
the gravitational degrees of freedom
\cite{Damour,Chirco,Bred1,Bred2,Thorne,Paddy}. This correspondence
allows the connection between macroscopic and microscopic physics
through the study of the statistical properties of the fluid on the
horizon of the black hole. For instance, the fluid on the horizon of a
Schwarzschild black hole can be modeled as a relativistic Bose gas
with all its degrees of freedom in the lowest energy level, allowing
the fluid to be in a condensed state \cite{Shanki1,Shanki2}.  The
collective behavior of the microscopic degrees of freedom, described
within the Landau-Ginzburg mean field theory of phase transitions, lead
to the Bekenstein-Hawking entropy in Einstein gravity
\cite{Shanki2,Shanki3} (see also \cite{Dvali}).  If the fluid-gravity
correspondence is a generic feature of any horizon, it is imperative
to see whether the mean field theory approach can be extended to
higher derivative gravity theories that are relevant in large
curvature limit where the Einstein-Hilbert action is not well suited.

Lanczos-Lovelock (LL) gravity is a generalization to Einstein 
gravity that is consistent with having no more than second order 
time derivatives in the equations of motion \cite{Lanc,Lov}, and is 
free from ghosts when perturbed around flat spacetime. The higher 
order terms in LL actions
represent high-energy corrections to Einstein gravity. In particular
the second order term known as the Gauss-Bonnet term, that is of 
interest in this work, appears in low
energy effective actions in string theories \cite{Zbach}, where the
Einstein-Hilbert term arises as the lowest order curvature term in the
action that is relevant at low energy scales. The black hole solution
related to the second order LL gravity, also known as
Einstein-Gauss-Bonnet (EGB) gravity, was found by Boulware and Deser
\cite{BoulDes} and a thermodynamic analysis of general LL gravity was
first done in \cite{Myers} (see also \cite{Paddy1,Paddy2}).

One of the main questions we ask here is the following: Is there a
statistical mechanical description for the fluid living on the BD
event horizon?  As mentioned earlier, such a description was recently
provided for Einstein gravity within mean field theory, where the 
fluid is modeled as a
Bose-Einstein condensate near the critical point \cite{Shanki2}. In
this work, we extend the statistical analysis to BD fluid
\cite{Jacob,Kolekar,Paddy1,Paddy2}.  As BD fluid contains high energy
corrections, one may expect that the mean field theory description can
not be extended naturally.  We construct an explicit model of the BD
fluid here where two phases coexist, a Bose-Einstein condensate and a
non-condensed (normal) phase. As we shall see, the condensed phase
corresponds to the Einstein-Hilbert term in the action whereas the
non-condensed phase arises due to the presence of the Gauss-Bonnet
term.

Before we go for the statistical analysis of BD fluid, we note that in
the literature there exists an ambiguity in the definition of 
entropy of the BD horizon fluid. Specifically, it has been argued that
the BD fluid entropy is less than the Wald entropy corresponding to
the BD black hole \cite{Kolekar,Paddy2}. We identify the root cause of
this ambiguity and show that this arises from assuming a linear
relation between free energy and the volume of the fluid.  We provide
a consistent thermodynamic framework and show that the entropy of the
fluid is given by the Wald entropy for the BD black hole, and forms 
the basis for further statistical analysis. 

We list below the steps followed to obtain the macroscopic quantities 
associated with the horizon-fluid from which we obtain a statistical mechanical 
description for BD fluid: 
\begin{enumerate}
 \item Using the expressions for pressure (\ref{Press}) and temperature (\ref{Temp})
 together with the free energy representation of the first law of thermodynamics (\ref{FL}), we define the
 thermodynamic Potential  ($\Omega$) or free energy   (\ref{WOmega}) corresponding to the horizon-fluid.
\item From the free energy (\ref{WOmega}), we obtain the entropy of the horizon-fluid. As mentioned above, 
there is an ambiguity in the literature regarding the value of this entropy~\cite{Kolekar,Paddy2}.  In Sec. 
(\ref{sec:entropyWald}), we explicitly show that the entropy of the fluid comes out to be identical to the 
Wald entropy when one takes into account that the free energy $\Omega$ is a non-linear function of  $A$. 
\item From the expression of free energy (\ref{WOmega}), we obtain the 
energy of the  horizon-fluid (\ref{Komar2}) using an integral of the form $\int PdV$. It is important to note 
that this definition of the energy of the fluid 
is different from the definition of Komar mass for the BD black hole although it matches with 
the accepted value of  energy (with no cosmological constant) given in the literature~\cite{Myers,Paddy1}. 
\end{enumerate}

The paper is organized as follows: In section II we resolve the
existing ambiguity in the expression for the entropy of the BD fluid,
which differs from the Wald entropy of the BD black hole
\cite{Kolekar,Paddy1,Paddy2}. We explicitly show how to obtain the entropy for
the BD fluid that leads to a value identical to the Wald entropy in a
consistent manner.  In section III, we apply the mean field theory for
the BD fluid and find that it shows the coexistence of two, condensate
and non-condensate, phases.  Under general assumptions, we provide a
statistical description of the non-condensed part of the fluid up to
first order in the coupling constant parameter $( \lambda )$ in the
Gauss-Bonnet term.  In section IV we conclude and discuss the
implications of our results. In appendix A, we generalize the
procedure followed in \cite{Shanki2} for the higher dimensional
Schwarzschild black hole and show that the correct value for
Bekenstein-Hawking entropy is also obtained as the difference of two
phases near a critical transition point. In appendix B we obtain the
generalized dispersion for
arbitrary dimensions.  Finally in appendix C, we deduce that a
dispersion relation with a linear term in momentum, namely a term
related to phonon modes, can also be present but is unimportant for
large values of momentum.
       
\section{Entropy of the Einstein-Gauss-Bonnet fluid}

The Lanczos-Lovelock Lagrangian, in a $D-$dimensional spacetime, is
the sum of particular Lagrangian densities, where each term is
characterized by a coupling constant $\lambda_m$ and each one of these
terms is a special contraction with products of completely
antisymmetric tensors $\delta^{a_1b_1...a_mb_m}_{c_1d_1...c_md_m}$
with Riemann tensors $R^{c_1d_1...c_md_m}_{a_1b_1...a_mb_m}$
\cite{Lanc,Lov}.  The first, second and third terms in the action
correspond to the cosmological constant, Einstein-Hilbert and
Gauss-Bonnet Lagrangian densities respectively. We will restrict
ourselves to second order LL Lagrangian with zero cosmological
constant and from now on, we refer $\lambda$ to the coupling constant
in the Gauss-Bonnet term, the action is given by

\begin{equation}
S = \int{d^D x \sqrt{-g} \left[ \frac{1}{16\pi} (R + \lambda \mathcal{L}_{GB}) \right]}. \label{Action}
\end{equation} 

\noindent where $\mathcal{L}_{GB} = R^2-4R^{\mu \nu}R_{\mu \nu} +
R^{\mu \nu \alpha \beta}R_{\mu \nu \alpha \beta}$ and $G$ is set to
unity. The spherically symmetric black hole corresponding to this
action was found in Ref. \cite{BoulDes} and is referred as BD black
hole.
  
As mentioned in the Introduction, the equations of motion of Einstein
gravity when projected to the surface horizon of its black hole give
rise to Navier-Stokes equations \cite{Damour}.  This projection has
been done for the equations of motion followed by Eq. (\ref{Action})
and for particular background geometries like the Boulware-Deser
\cite{Jacob}. We refer to the fluid in this particular background
geometry as BD fluid. The transport coefficients related to pure
LL terms in Eq. (\ref{Action}) have been found in
Ref. \cite{Kolekar}. The projection from the equations of motion was
made using a limiting process from an stretched horizon to the true
horizon and a detailed discussion of this process and its validity is
given in Refs. \cite{Jacob,Kolekar,Wilczek}.  The pressure and
temperature for the $D-$dimensional BD fluid are given by
\cite{Jacob,Kolekar}
\begin{equation}
  P = \frac{(D-3)}{16\pi a} + \frac{\bar{\alpha}(D-5)}{16\pi a^3}, \label{Press}
\end{equation}
        
\begin{equation}
T = \frac{(D-3)a^2 + (D-5)\bar{\alpha}}{4\pi a^3 + 8\pi \bar{\alpha}a}. \label{Temp}
\end{equation}

\noindent where $a$ is the horizon radius of the BD black hole, and
$\bar{\alpha} = \lambda(D-3)(D-4)$.  Now we turn our attention to the
relation between the pressure and entropy of the fluid on the
horizon. The prescription to calculate the entropy for a general
diffeomorphism invariant theory of gravity for a killing horizon was
given by Wald \cite{Wald}. Following this prescription, the Wald
entropy for the BD black hole is given by \cite{Paddy1,Ron}

\begin{equation}
  S_W = \frac{A}{4} + \lambda \frac{(D-2)(D-3)(A_{D-2})^{\frac{2}{D-2}}}{2}A^{\frac{D-4}{D-2}}, \label{WaldEn}
\end{equation}

\noindent where $A_{D-2}$ is the area of the unit $(D-2)$ dimensional
sphere.  While Wald entropy is uniquely defined, it seems that the
entropy for the BD fluid is not defined uniquely
\cite{Kolekar,Paddy2}.  In particular, using
Eqs. (\ref{Press}),(\ref{Temp}), the pressure can be expressed as

\begin{equation}
  P = \frac{T}{4} + \frac{\bar{\alpha}(A_{D-2})^{\frac{2}{D-2}}}{2}TA^{-\frac{2}{D-2}},    \label{PrAT}
\end{equation}

\noindent where this relation can be understood as an equation of
state for the pressure as a function of temperature and area.  From
this relation we can easily see that the pressure for the
Schwarzschild fluid is recovered for $\lambda = 0$ \cite{Shanki2}.
Eq. (\ref{PrAT}) can be rewritten to find another interesting relation
between pressure and entropy:

\begin{equation}
  \frac{PA}{T} = \frac{A}{4} + \frac{\bar{\alpha}(A_{D-2})^{\frac{2}{D-2}}}{2}A^{\frac{D-4}{D-2}}. \label{EqState}
\end{equation}

\noindent Comparing with Wald entropy Eq. (\ref{WaldEn}), we can write
Eq. (\ref{EqState}) in the following way

\begin{equation}
\frac{PA}{T} = S_W^1 + \frac{(D-4)}{(D-2)}S_W^2, \label{EqState2}
\end{equation}

\noindent where the upper index designates the corresponding
contribution to Wald entropy of individual LL terms in
Eq. (\ref{Action}), i. e.

\begin{align}
& S^1_W = \frac{A}{4}, \\ \nonumber
& S^2_W = \lambda \frac{(D-2)(D-3)(A_{D-2})^{\frac{2}{D-2}}A^{\frac{D-4}{D-2}}}{2}.
\end{align}

\noindent The relation (\ref{EqState2}), for the general LL
Lagrangian was found in \cite{Kolekar,Paddy2}, and it is the entropy
of the BD fluid. This entropy is explicitly given by

\begin{equation}
S_{F} = \frac{A}{4} + \lambda \frac{(D-3)(D-4)(A_{D-2})^{\frac{2}{D-2}}}{2}A^{\frac{D-4}{D-2}}, \label{MPEn}
\end{equation}

\noindent and differs from Wald entropy Eq. (\ref{WaldEn}) by a
numerical factor in the second term. It is important to note that
$S_{F}$ is always less than $S_{W}$ $(S_{F} < S_{W})$.  In the
remainder of this section we first point out the reason for the
mismatch and show a consistent way to relate the thermodynamic
quantities of the BD fluid. In this approach, the entropy of the BD
fluid coincides with the Wald entropy for the BD black hole.

\subsection{Thermodynamics of BD fluid}
\label{sec:entropyWald}

Here we point out the origin of the discrepancy between the two
entropies, namely the Wald entropy for the BD black hole
Eq. (\ref{WaldEn}), and the entropy for the BD fluid given by
Eq. (\ref{MPEn}). The main observation is that Eq. ({\ref{PrAT}})
is an equation of state for the pressure, that should be derived from
a thermodynamic potential as a fundamental relation, namely the free
energy $\Omega (A,T)$ satisfying the differential first law of
thermodynamics \cite{note1}

\begin{equation}
d\Omega = -SdT - PdA, \label{FL}
\end{equation}     

\begin{equation}
S = -\frac{\partial \Omega}{\partial T} \bigg \vert_{A}, ~~~~~P = -\frac{\partial \Omega}{\partial A} \bigg \vert_{T}. \label{PressEntro}
\end{equation}

\noindent At this point it is important to remark that if we attempt
to construct the free energy from the expression, $\Omega(A,T) = -PA$
with the pressure given by Eq. (\ref{PrAT}) we see that the
thermodynamic equations (\ref{PressEntro}) give the entropy as in
Eq. (\ref{MPEn}) and an incorrect expression for the pressure. The
reason why the last construction fails is because the energy $U(A,T)$
is not a homogeneous function of order one in its extensive variables
and, as a consequence of this, the Euler relation $U = TS-PA$ does not
hold, which is a necessary requirement for the free energy to be given
by $\Omega = -PA$ \cite{Thermo}. This is a particular feature of the LL
thermodynamic relations. Therefore, we need to construct the free
energy $\Omega(A,T)$ from the differential relation (\ref{FL}). Using
the input, that is the pressure (\ref{PrAT}), and
imposing that the entropy at the origin vanishes we get the following
free energy
 
\begin{equation}
\Omega = -\frac{TA}{4} - \lambda \frac{(D-2)(D-3)(A_{D-2})^{\frac{2}{D-2}}}{2}TA^{\frac{D-4}{D-2}}. \label{WOmega}
\end{equation}

\noindent The equations of state related to this $\Omega (A,T)$ are
the correct expression for pressure Eq. (\ref{PrAT}) and Wald entropy
Eq. (\ref{WaldEn}) for the (BD) black hole \cite{Myers,Ron,Paddy1}. In
this way, the thermodynamic relations for the BD fluid are seen to be consistent
and the problem of the ambiguity in entropy disappears, i.e. the
entropy of the BD fluid is the same as Wald entropy for the BD black
hole of EGB gravity.

\noindent We also see that it follows directly from the equations of
state that $\frac{\partial S}{\partial A} = \frac{P}{T}$, and this
quantity depends solely on the area $A$ which means that

\begin{equation}
dU = TdS -PdA = 0,
\end{equation}  

\noindent This relation is consistent with the corresponding BD black
hole constraints as one parameter system depending on the area of its
horizon $A$ or the black hole radius $a$. The energy corresponding to the 
horizon-fluid can be obtained from the free energy $\Omega (A,T)$, i. e.,
\begin{equation}
  E = \int PdA = \int{\left( -\frac{\partial \Omega (A,T)}{\partial A} \bigg \vert_{T} \right)} dA. \label{Komar2}
\end{equation}
\noindent Substituting  Eq.~(\ref{WOmega}) for the free energy in the above expression leads to 
\begin{align}
E = & \frac{(D-2)(A_{D-2})^{\frac{1}{D-2}}A^{\frac{D-3}{D-2}}}{16\pi} \label{KomarE} \\ \nonumber 
& + \lambda \frac{(D-2)(D-3)(D-4)(A_{D-2})^{\frac{3}{D-2}}A^{\frac{D-5}{D-2}}}{16\pi} \, . 
\end{align}
This is one of the key results of this work and we would like to stress the following points: 
First, the thermodynamic quantities are obtained from the free energy and are 
computed from an independent construction using the properties of the BD fluid. 
More importantly, the thermodynamic properties of the BD black hole calculated using 
the Euclidean action approach \cite{Myers}, and the procedure followed here are independent.  
Second, although the thermodynamic quantities are calculated using a different approach, they 
match with the correct accepted values for the BD black hole~\cite{Myers,Paddy1,Ron,Katz3}. The matching 
of the macroscopic quantities from geometric and thermodynamic routes reinforces the 
correspondence between the fluid and gravity on the horizon. Third, there are several inequivalent 
definitions of mass in general relativity. The mass or the ``energy" of a black hole is also not free from 
this problem. For global conditions like staticity and asymptotically flatness (zero cosmological constant), 
the Komar integral gives the correct mass parameter, at least for Schwarzschild black holes in 4 
dimensions~\cite{Beig,Ashtekar,Laszlo}. 
However,  the definition of Komar energy for the  Kerr black hole \cite{Katz1,Katz2} has an anomalous factor $2$. 
The value of energy from our calculation matches the correct value of energy for the BD black hole \cite{Myers,Paddy1,Katz3}, 
which is not based on Komar-like calculations.

\section{Statistical modeling of the BD horizon fluid}

In this section we turn our attention to the statistical modeling of
the BD fluid. 

Following the same procedure as in \cite{Shanki2}, we show in appendix
A that Bekenstein-Hawking entropy arises as the difference in the
entropy between two phases near a critical point for the higher
dimensional Schwarzschild black hole. This means that the mean field
theory is well suited to describe the Schwarzschild fluid near the
critical point and all its degrees of freedom are in the ground
state. This will correspond to the condensed phase of the BD fluid.

Here we note the fact that EGB gravity is understood as a gravity
theory containing high energy corrections or equivalently
contributions coming from the physics at a much shorter length scale
compared to the scale where Einstein gravity is a good
approximation \cite{Zbach}, It is well known, that mean field theory
description of the BEC includes only the long wavelength physics and
is independent of the physics at smaller length scales. Hence it is
expected that the part of the fluid corresponding to the GB term
cannot be described by a BEC. In fact, one encounters problems if one
tries to model GB horizon fluid by a mean field theory. Accordingly,
we shall treat the part of the horizon fluid corresponding to the GB
term as being in a non-condensed or normal phase.  We are interested
in understanding how these high energy contributions manifest in the
fluid description of the BD black hole.  We can separate what we call
now the total free energy $\Omega$ into two parts $\Omega_C$ and
$\Omega_{N}$ corresponding to the condensed and non-condensed parts
respectively. They are given by,

\begin{align}
  & \Omega_C = -\frac{TA}{4}, \label{Omegas} \\ \nonumber & \Omega_{N}
  = -\lambda
  \frac{(D-2)(D-3)(A_{D-2})^{\frac{2}{D-2}}}{2}TA^{\frac{D-4}{D-2}}.
\end{align}

\noindent The free energy $\Omega_C$ is given by the mean field theory
expansion (see details in appendix A),

\begin{equation}
\Omega_C = \Phi_0 + a(T-T_c)\eta^2 + B\eta^4.
\end{equation}

\noindent The following points are in order regarding the above
expression: First, it tells us that the properties of the condensed
part can be described by a mean field theory. This part does not
explicitly depend on $\lambda$ and it matches with the free energy of
Einstein gravity. Second, the free energy part that explicitly depends
on $\lambda$ can not be described as a relativistic Bose gas. In
hindsight, this is not surprising as Gauss-Bonnet gravity corresponds
to high-energy corrections. In the fluid description, this translates
to the non-condensed part of the fluid that comprises of the
excitations of the medium. In other words, the statistical description
of the non-condensate part can provide us with information about the
high energy features of the gravity theory.
   
\subsection{Thermodynamic quantities characterizing the non-condensed  phase of the fluid}

Having identified the free energy corresponding to the non-condensed
part of the fluid, we provide the statistical mechanical description
for this non-condensate phase characterized by $\Omega_N$ in
Eq. (\ref{Omegas}).  To achieve that, first we obtain the
thermodynamic relations for the non-condensed part of the fluid. The
free energy $\Omega_{N}$ (\ref{Omegas}) satisfies the thermodynamic
relation,

\begin{equation}
d \Omega_{N} = -S_{N}dT-P_{N}dA.
\end{equation}
 
\noindent The thermodynamic quantities for the non-condensed fluid
that follow from this expression are given by,

\begin{align}
  & S_{N} = \lambda
  \frac{(D-2)(D-3)(A_{D-2})^{\frac{2}{D-2}}}{2}A^{\frac{D-4}{D-2}}, \label{NCExp}
  \\ \nonumber & P_{N} = \lambda
  \frac{(D-3)(D-4)(A_{D-2})^{\frac{2}{D-2}}}{2}TA^{-\frac{2}{D-2}}, \\
  \nonumber & E_{N} = \lambda
  \frac{(D-2)(D-3)(D-4)(A_{D-2})^{\frac{3}{D-2}}A^{\frac{D-5}{D-2}}}{16\pi},
\end{align}

\noindent where,

\begin{align}
& S = S_{C} + S_{N}, \label{Sum} \\ \nonumber
& P = P_{C} + P_{N}, \\ \nonumber
& \Omega = \Omega_{C} + \Omega_{N}, \\ \nonumber 
& E = E_{C} + E_{N}.
\end{align}

\noindent The expressions for $S, P, \Omega$ and $E$ are given by
Eqs. (\ref{WaldEn}), (\ref{PrAT}), (\ref{WOmega}), and
Eq. (\ref{KomarE}) respectively.

\subsection{Statistical modeling of the non-condensed fluid}

Since we are working with the fluid description of the BD horizon, the
degrees of freedom in the non-condensed phase can be viewed as the
excitations of the medium (i.e. the fluid) due to the vibrational
modes.  To go about with the statistical modeling of the non-condensed
phase of the fluid, we consider it to be a weakly interacting gas. The
following three physical assumptions can then be made: (i) As we have
shown in the last section, we are dealing with a Bose-like fluid with
a fraction of its degrees of freedom in the condensate state, we
assume a Bose-Einstein probability distribution for the vibrational
modes in the fluid as a function of energy in the grand canonical
ensemble with zero chemical potential.  (ii) We will assume that the
density of states is proportional to the volume accessible to the
fluid, in this case, the area of the horizon. (iii) Finally, we will
assume a power law for the density of states, i.e. the dispersion
relation between energy and momentum is given by, $\epsilon_k =
ck^{\sigma}$, where, $\sigma$ is the spectral index \cite{BCD1,BCD2}.
This relation can be recovered knowing the density of states
$n(\epsilon)$,
\begin{equation}
  \frac{d \Sigma (\epsilon) d\epsilon}{d\epsilon} = n(\epsilon)d\epsilon. \label{Density} 
\end{equation}   

\noindent where, $\Sigma (\epsilon)$ is the number of microstates in
this infinitesimal interval. After taking into account the last two
assumptions, the density of states takes the form,

\begin{equation}
n(\epsilon)d\epsilon = aA\epsilon^{\gamma} d\epsilon, \label{Density2} 
\end{equation}   

\noindent where, $a$ and $\gamma$ are constants to be determined by
the constraints of the model. Note that $\sigma$ and $\gamma$ are
different.  Following this, the logarithm of the partition function
$\ln{\mathcal{Q}}$ is given by,

\begin{equation}
  \ln{\mathcal{Q}} = -aA\int_0^{\infty}\epsilon^{\gamma}\ln{(1-e^{-\beta \epsilon})} d\epsilon, \label{LogQ}
\end{equation}
\noindent and the energy is given by

\begin{equation}
  \bar{E} = -\frac{\partial (\ln{\mathcal{Q}})}{\partial \beta} = aA\Gamma(\gamma + 2)\zeta(\gamma + 2)(k_{B}T)^{\gamma + 2} \label{EnGen}
\end{equation}

\noindent where, $\Gamma(x)$ and $\zeta(x)$ are the Euler gamma
function and Riemann zeta function respectively. Let us focus on $D=6$
dimensions.  The thermodynamic properties are given by,

\begin{align}
& E_N =  \frac{3 \lambda (A_4)^{\frac{3}{4}}A^{\frac{1}{4}}}{2\pi}, \label{NCQ6} \\ \nonumber
& P_N = 3 \lambda (A_{4})^{\frac{1}{2}}TA^{-\frac{1}{2}}, \\ \nonumber
& S_N = 6 \lambda (A_{4})^{\frac{1}{2}}A^{\frac{1}{2}}. \\ \nonumber 
\end{align}

\noindent We see that all the relevant thermodynamic quantities are
functions of $T$ and $A$, but $T(A)$ is function of $A$ itself, so we
have have a one parameter system. Therefore, we can express the
thermodynamic quantities as a function of one of this two variables
that we choose here to be the temperature. For this purpose we rewrite
Eq. (\ref{Temp}) as,

\begin{equation}
  T(A) = \frac{3(A_{4})^{\frac{1}{4}}A^{\frac{1}{2}} + 6 \lambda (A_{4})^{\frac{3}{4}}}{4\pi A^{\frac{3}{4}}+48 \pi \lambda (A_{4})^{\frac{1}{2}}A^{\frac{1}{4}}}, \label{1}
\end{equation}

\noindent and we use \eqref{1} to express area as a function of
temperature $A(T)$ to first order in $\lambda$. Substituting back on
Eq. (\ref{NCQ6}) we get,

\begin{eqnarray}
& \frac{E_N}{A} = \frac{32\lambda \pi^2 T^3}{9}, \label{Thermo6} \\ \nonumber \\ \nonumber
& P_N = \frac{16 \lambda \pi^2 T^3}{3}, \\ \nonumber
& S_N = \frac{27\lambda(A_4)}{8\pi^2 T^2}.
\end{eqnarray}

\noindent Matching the thermodynamic energy density
Eq. (\ref{Thermo6}) with the statistical energy density coming from
Eq. (\ref{EnGen}), $a$ and $\gamma$ are given by,

\begin{equation}
a= \frac{16\lambda \pi^2}{9\zeta(3)k_{B}^3},~~~~~~~~\gamma = 1. \label{agamma}
\end{equation}

\noindent Using the values of $a$ and $\gamma$ in Eq. (\ref{LogQ}) we
get,

\begin{equation}
\ln{\mathcal{Q}} = \frac{16\lambda \pi^2 A T^2}{9 k_{B}}.
\end{equation}
 
\noindent The statistical equations of state for the pressure and
entropy are

\begin{equation}
\bar{P} = \frac{\partial (k_{B}T\ln{\mathcal{Q}})}{\partial A} \bigg 
\vert_{T},
~~~~\bar{S} = \frac{\partial (k_{B}T\ln{\mathcal{Q}})}{\partial T} \bigg \vert_{A}.
\end{equation}

\noindent As partial derivations are involved in finding these
equations of state, it is important to note that
$k_{B}T\ln{\mathcal{Q}}$ is defined up to some functions of pressure
and temperature. Making use of this freedom, we shall match the
expressions for the variables from the thermodynamics and
statistical mechanics. To this end, we define,

\begin{equation}
  \bar{\Omega} (A,T) = k_{B}T\ln{\mathcal{Q}} + \frac{6\lambda (A_4)^{\frac{3}{4}} A^{\frac{1}{4}}}{\pi} - \frac{27\lambda (A_4) }{16\pi^2 T}. \label{SkTQ}
\end{equation}

\noindent An expression from which we get the statistical quantities

\begin{align}
  \bar{P} = \frac{\partial \bar{\Omega}}{\partial A} \bigg \vert_{T},
  ~~~\bar{S} = \frac{\partial \bar{\Omega}}{\partial T} \bigg
  \vert_{A} \label{2}
\end{align}

\noindent Rewriting the statistical mechanical quantities obtained
from \eqref{2} in terms of $T$ we get,

\begin{equation}
  \bar{P} = \frac{16\lambda \pi^2 T^3}{3},~~~~~~ \bar{S} = \frac{27\lambda(A_4)}{8\pi^2 T^2}.
\end{equation}

\noindent It is important to note that this matches with the quantities (\ref{Thermo6}) as obtained from the thermodynamics of BD fluid. 

Now that we have a complete statistical picture of the non-condensed
part of the fluid, we can gain more insight into the microscopic
features of the BD fluid by looking at the dispersion relation
satisfied by the vibrational modes. To this end, we look at the
differential relation involving the density of states $n(\epsilon)$,

\begin{equation}
n(\epsilon)d\epsilon = g(p)dp, \label{Disp}
\end{equation} 

\noindent using (\ref{GenDisp}) (with $\hbar = 1$) for the
$D=6$-dimensional space-time (4-dimensional fluid), $g(p)dp$ is given
by,

\begin{equation}
g(p)dp = \frac{Ap^3}{8 \pi^2}dp, \label{Dispp}
\end{equation}

\noindent and using Eq. (\ref{Density2}) with the value for $a$ given
in Eq. (\ref{agamma}), the non-condensed fluid satisfies the following
dispersion relation,

\begin{equation}
\epsilon = \frac{3\sqrt{\zeta(3)}k_{B}^{3/2}}{16\pi^2 \sqrt{\lambda}}p^2. \label{Disp6}
\end{equation}

\noindent The following points are note worthy regarding the above
results: First, the non-condensed phase satisfies the dispersion
relation of a non-relativistic gas with $\lambda$ playing the role of
mass.  When $\lambda << 1$, which corresponds to the approximation we
used, the mass of the particles is small and this dispersion relation
corresponds to the excitation of low energy modes.  Phonon modes are
present but they are not important for large values of momentum (see
appendix C).  Second, at the leading order in $\lambda$, the above
analysis can be extended to any dimensions. It can be seen that, for
any dimension, the energy constraint can be satisfied for $\gamma = 1$
and the dispersion relation will depend on the space-time dimensions.
These generalized dispersion relations arise naturally in higher
dimensional Bose-Einstein condensates \cite{BCD1,BCD2} and we can see
generally that the condensation property is satisfied for higher
dimensional fluids like the ones encountered here.

\section{Conclusions}                  
            
The paradigm that gravity is emergent is supported by quite a few
independent lines of evidences. The connection of the gravitational
dynamics with the Navier-Stokes equation is definitely one of the most
striking among these. However, except for the Fluid-Gravity duality
that emerges from the AdS-CFT correspondence, it is not known whether
there is an underlying microscopic theory corresponding to the fluid
description of the dynamics of horizons.  Though the AdS-CFT correspondence is
restricted to certain classes of black hole spacetimes, still it
is possible to glean a lot of insight about the microscopic theory of
gravity via Fluid-Gravity duality. It is interesting to find out what
the microscopic theory underlying a fluid description of the event
horizon of a black hole is in a gravity theory. For the Schwarzschild
black hole spacetime, a solution of the Einstein theory, this study
has already been performed~\cite{Shanki1}. Here we have extended 
this approach to include black holes in Lanczos-Lovelock gravity.

In this work we modeled the horizon fluid of BD black hole as a fluid
with two coexisting phases; a BEC phase corresponding to the Einstein
gravity and a non-condensed phase that corresponds to the Gauss-Bonnet 
term. Our analysis has two parts: (i) Purely thermodynamic and (ii) 
Statistical mechanical modeling of the fluid. 

In the thermodynamic aspect of our work, we provided  a consistent way 
to derive the equations of state for the general $D$-dimensional horizon-fluid 
in Lovelock theory and showed that the entropy of this fluid coincides with the 
Wald entropy~\cite{Wald,Ron}. Our prescription resolves the ambiguity 
existing in the literature in the definition of the entropy for the black hole 
solution of EGB gravity in its fluid description~\cite{Kolekar,Paddy2}, and 
establish a solid basis for the thermodynamics of the BD fluid.

On the statistical mechanical part, we generalized the results found
in \cite{Shanki2,Shanki3} for the higher dimensional Schwarzschild
black hole and using the mean field theory analysis, we find that
the BD fluid shows the coexistence of a condensed phase and a
non-condensed (normal) phase.  Finally, to have a better insight in
the nature of the non-condensed fluid, we provided detailed statistical 
analysis.  Under general assumptions, we found
a consistent statistical model of the non-condensed part of the BD
fluid.  In the approximation used, we find that the normal fluid
behaves like a non-relativistic fluid with a general dispersion
relation given by Eq. (\ref{GenDisp}).  The 6-dimensional model is
treated in detail where the coupling constant $\lambda$ plays the
role of mass for the low energy modes of the excited part.  In the 
limit of $\lambda \to 0$, we recovered Schwarzschild fluid condensate.

Our work implies that the counting of the DOF on the black hole event horizon in Gauss-Bonnet gravity is 
greater than 
that for the Einstein theory of gravity. This is consistent with results of Brustein and Medved~\cite{Brustein}, where 
the authors have shown that the Lovelock theory of gravity can be effectively described as Einstein gravity 
coupled to a 2-form gauge field.  This is reminiscent of the $f(R)$ gravity in a conformally transformed frame, where it can be 
described as Einstein gravity coupled to a scalar field. Also from the formula for Wald entropy for a black 
hole in a Lovelock theory \cite{Paddy1,Myers,Ron}, we see that the entropy is greater if the 
Gauss-Bonnet coupling term is non-zero. While the first point  evidence that the Lovelock theory of 
gravity has extra degrees of  freedom compared to the Einstein theory of gravity, the Wald entropy 
formula shows that black hole entropy in Gauss-Bonnet theory is greater than the entropy of the 
black holes in the Einstein theory of gravity { (for the black holes having the same area in both 
theories)}.  Even though in 4-dimensions, Gauss-Bonnet term is topological, it gives a non-vanishing constant 
term to the Wald entropy.  In other words, in four dimensions, the entropies of the black hole in the two theories 
differ by a constant term. This also seems to indicate that if the Wald entropy denotes the number of 
microstates of a black hole,  then probably that number is greater for a black hole in Lovelock gravity.  
This is true even if the coupling constant is small. To our understanding, the reason has not been understood completely.

At this point, it is also interesting to note that the Boulware-Deser black hole solution lies in one branch. 
The solution on the other branch has a naked singularity. The approach pioneered by Damour works requires 
event horizon, i.e. a null horizon generated by a Killing vector. Since horizon fluid cannot be defined for 
the other branch of solutions, hence our approach cannot be extended to provide a fluid description of such 
solutions. 

In this work, we have considered horizon-fluid of Boulware-Deser 
black hole which is an asymptotically flat. Our aim is extend the 
analysis for asymptotically A(dS) space-times. In the case of Einsteinian 
gravity, we showed that a negative cosmological constant acts 
like an external magnetic field that induces order in the system 
leading to the appearance of a tri-critical point in the phase diagram~\cite{Shanki2}.
In the case of 5-D Gauss-Bonnet gravity with (positive or negative) cosmological 
term, $\lambda = 1/4$ (in Geometric units) has a critical value and the symmetry 
enhances to the full SO(4, 2) group and the 5-D Gauss-Bonnet gravity action 
with cosmological term is Chern-Simons Lagrangian for the AdS group~\cite{Chamseddine}
It will be interesting to obtain a fluid description for the critical system as it will 
be non-perturbative. We hope to address this elsewhere.

The coexistence of two phases in our model is a feature that is common
to the two fluid model of Superfluidity as well. Given the context, it
is natural to ask the question, whether a two fluid model can be
developed that describes the dynamics of the BD fluid. If possible, we
might be able to relate the physics of the BD fluid to the physics of
superfluidity where the coexistence of two phases occurs naturally.
The two fluid models also exhibit some other interesting
properties \cite{SFluids} and one could check whether those features
are seen in this case also. We hope to address these questions
elsewhere.

\appendix

\section{Higher dimensional Schwarzschild black hole entropy from criticality}

Following the same procedure as in \cite{Shanki2}, we show that for
general higher dimensional Schwarzschild black hole,
Bekenstein-Hawking entropy arises as the entropy difference between
two phases near a critical point.  We start from the generalized
expressions Eqs. (\ref{Temp}), (\ref{KomarE}) with $\lambda = 0$,
being the relevant ones the Komar energy and temperature appearing in
the order parameter $\eta$ in the statistical field (free energy)
expansion \cite{Shanki1,Landau}. The energy and temperature for
$\lambda = 0$ in terms of area $A = A_{D-2}a^{D-2}$ are given by

\begin{align}
& E = \frac{(D-2)(A_{D-2})^{\frac{1}{D-2}}A^{\frac{D-3}{D-2}}}{16 \pi}, \label{ED0} \\ \nonumber    
& T = \frac{(D-3)(A_{D-2})^{\frac{1}{D-2}}}{4\pi A^{\frac{1}{D-2}}}.
\end{align}

\noindent Now, we can construct the quantity $N(A) = E/ \alpha T$, we
get

\begin{align}
N(A) = \frac{(D-2)A}{4\alpha(D-3)}. \label{ND0}
\end{align}

\noindent The form of the constraint $N(A)$ needed to construct the
order parameter $\eta$ was justified from the micro canonical point of
view for the Schwarzschild black hole fluid ($D=4$) in
\cite{Shanki1,Shanki2}, and this statement can be generalized to the
$D-$dimensional case in which we are interested now and the same
functional form of $N(A)$ is found. The order parameter $\eta^2$ is
given by $\eta^2 = \kappa N(A)$. We can also see from Eq. (\ref{PrAT})
that in this higher dimensional case, it is also satisfied that the
form of the equation of sate is $P = T/4$.  We can use the
thermodynamic potential for this case, Eq. (\ref{WOmega}) given by
$\Omega (A,T) = -TA/4$ and the mean field theory expansion is
\cite{Landau,Shanki1}
 
 \begin{equation}
 -\frac{TA}{4} = \Phi_0 + a(T-T_c)\eta^2 + B\eta^4,
% & -\frac{TA}{4} = \Phi_0 + a(T-T_c)\left[ \frac{\kappa(D-2)A}{4(D-3)\beta}  \right] + B\left[ \frac{\kappa(D-2)A}{4(D-3)\beta} \right]^2,
 \end{equation}

 \noindent when matching coefficients on both sides of the equation we
 get the value of $a$, $a = -\frac{(D-3)\alpha}{(D-2)\kappa}$, and the
 value of $\eta$ of the extremum of the statistical field is obtained
 from $\partial \Phi/ \partial \eta = 0$, this is

\begin{equation}
\eta^2 = -\frac{a(T-T_c)}{2B}, 
\end{equation} 

\noindent where using $\eta^2 = \frac{\kappa(D-2)A}{4(D-3)\alpha}$ we
get

\begin{equation}
\frac{(T-T_c)}{2B} = \frac{(D-2)^2\kappa^2}{4(D-3)\alpha^2}A.
\end{equation} 

\noindent Finally, we find the entropy, $\triangle S = - \partial \Phi
/ \partial T$

\begin{equation}
\triangle S = -\frac{\partial \Phi}{\partial T} = -a\eta^2 = \frac{a^2(T-T_c)}{2B},  
\end{equation}

\noindent and substituting the previous values this gives

\begin{equation}
\triangle S = \frac{A}{4}.
\end{equation}

\noindent We find that this difference in entropy between the two
phases of the fluid near the critical point is in accordance with the
Bekenstein-Hawking entropy.

\section{General dispersion relation}

For a general $(D-2)$ dimensional fluid on the horizon of a
$D$-dimensional black hole, the number density of the states for the
vibrational modes of the fluid is given by,
\begin{equation}
\Sigma(p) = \frac{A}{h^{D-2}}\int d^{D-2} p = \frac{A}{h^{D-2}}\frac{\pi^{\frac{D-2}{2}}p^{D-2}}{\Gamma \left( \frac{D}{2}\right)}
\end{equation}  

\noindent and from this we get $g(p)$

\begin{equation}
\frac{d \Sigma(p)}{dp}dp = g(p)dp = \frac{(D-2)A\pi^{\frac{D-2}{2}}p^{D-3}}{h^{D-2}\Gamma \left( \frac{D}{2}\right)}
\end{equation}

\noindent and finally from the relation $n(\epsilon)d\epsilon =
g(p)dp$ we get the generalized dispersion relation,
\begin{equation}
\epsilon = \frac{p^{\frac{D-2}{2}}}{2^{\frac{D-3}{2}}\pi^{\frac{D-2}{4}}\hbar^{\frac{D-2}{2}} \sqrt{\Gamma (D/2)}\sqrt{a}} \label{GenDisp}
\end{equation}

\noindent where, the value of the constant $a$ depends on $\lambda$
and can be fixed using the energy constraint.

\section{Dispersion relation with a linear term}

Here we show that the leading term in the dispersion relation is
$\epsilon_0 (p) = cp^2$ [see Eq. (\ref{Disp6})], while the other terms
in the dispersion relation are subleading corrections.  Since we are
considering the excitations in a fluid medium, it is natural to
consider an UV cut-off. Then the total energy of the non-condensed
part of the fluid in terms of momentum is given by,
\begin{align}
\bar{E} = \int_0^{p_D} \frac{g(p) \epsilon (p)}{e^{\beta \epsilon(p)}-1}dp. \label{App1}
\end{align} 

\noindent where, $p_D$ is the cut-off. We now recall that the
thermodynamic expression for the energy is equated with the
statistical average assuming no cut-off in Eq. (\ref{EnGen}) (with
$\gamma = 1$) where $\epsilon (p)$ is given by $\epsilon_0 (p)$.  In
order to recover the average energy $\bar{E}$, $\epsilon (p)$ in
Eq. (\ref{App1}) has to be slightly different from $\epsilon_0
(p)$. We may express this as, $\epsilon (p) = \epsilon_0 (p) + \delta
\phi(p)$, where $\delta \phi(p)$ represents the change in the
dispersion relation which will be of the order of $(1/p_D)$.  So, the
relation (\ref{App1}) can be written as,

\begin{align}
  \int_0^{p_D} \frac{g(p) \epsilon (p)}{e^{\beta \epsilon(p)}-1}dp
  \rightarrow \int_0^{\infty} \frac{g(p) [\epsilon (p) - \delta \phi
    (p)]}{e^{\beta [\epsilon (p) - \delta
      \phi(p)]}-1}dp.  \label{App2}
\end{align} 

\noindent Expanding the right hand side of Eq. (\ref{App2}) up to the
first order in $1/p_D$ we get,

\begin{align}
  \bar{E} = & \int_0^{\infty} \frac{g[p(\epsilon)] \epsilon}{e^{\beta
      \epsilon}-1}\left( \frac{dp}{d \epsilon} \right)
  d\epsilon \label{Echange} \\ \nonumber & - \frac{1}{p_D} \left[
    \frac{g(p^{\prime})}{e^{\beta \epsilon (p^{\prime})}-1} - \beta
    \frac{g(p^{\prime})\epsilon (p^{\prime})e^{\beta \epsilon
        (p^{\prime})}}{[e^{\beta \epsilon (p^{\prime})}-1]^2} \right].
\end{align}

\noindent The explicit change in $\epsilon (p)$ is given by,
$\epsilon(p) = \epsilon_0 (p) + \frac{1}{p_D}[a_1p + a_2p^2 + a_3p^3 +
...]$, where, $a_1, a_2, a_3, ...$ are arbitrary constants and are
constrained by a consistency condition which will be derived below.
We invert the expression of $\epsilon(p)$ to express $p$ as a function
of $\epsilon$ perturbatively, i. e.

\begin{equation}
  p(\epsilon) = p_0 (\epsilon) + \frac{p_1(\epsilon)}{p_D} + \frac{p_2(\epsilon)}{p_D^2} + ... \label{p}
\end{equation}

\noindent where, $p_0 = \left(\epsilon/c \right)^{1/2}$. We find
$p(\epsilon)$ up to the coefficient $a_3$ and to first order in
$1/p_D$. The factor $g(p)$ for the six dimensional case is given by,
$g(p) = bp^3$, where, $b$ is a constant Eq. (\ref{Dispp}),  in
  this case $g[p(\epsilon)] = bp^3(\epsilon)$. Hence,
substituting $p(\epsilon)$ in Eq. (\ref{Echange}) up to first order in
$1/p_D$, we get,

\begin{align}
  \bar{E} = & \frac{b}{2c^2}\int_0^{\infty}{\frac{\epsilon^2
      d\epsilon}{e^{\beta \epsilon}-1}} - \frac{1}{p_D} \left \{
    \frac{9a_1b\sqrt{\pi}}{16(c\beta)^{5/2}}\zeta \left( \frac{5}{2}
    \right) \right\} \\ \nonumber 
& - \frac{1}{p_D} \left\{
    \frac{a_2b}{2(c\beta)^{3}}\zeta(3) +
    \frac{75a_3b\sqrt{\pi}}{32(c\beta)^{7/2}}\zeta \left( \frac{7}{2}
    \right) \right\} \\ \nonumber 
& - \frac{1}{p_D} \left[
    \frac{g(p^{\prime})}{e^{\beta \epsilon (p^{\prime})}-1} - \beta
    \frac{g(p^{\prime})\epsilon (p^{\prime})e^{\beta \epsilon
        (p^{\prime})}}{[e^{\beta \epsilon (p^{\prime})}-1]^2} \right].
\end{align}

\noindent We note that the first term on the right hand side is the
average energy $\bar{E}$ [see Eq. (\ref{EnGen})]. Using this, we find
the consistency condition,

\begin{align}
  & \left \{ \frac{9a_1b\sqrt{\pi}}{16(c\beta)^{5/2}}\zeta \left(
      \frac{5}{2} \right) + \frac{a_2b}{2(c\beta)^{3}}\zeta(3) +
    \frac{75a_3b\sqrt{\pi}}{32(c\beta)^{7/2}}\zeta \left( \frac{7}{2}
    \right) \right\} \\ \nonumber
  & = - \left[ \frac{g(p^{\prime})}{e^{\beta \epsilon (p^{\prime})}-1}
    - \beta \frac{g(p^{\prime})\epsilon (p^{\prime})e^{\beta \epsilon
        (p^{\prime})}}{[e^{\beta \epsilon (p^{\prime})}-1]^2}
  \right].  \label{Final}
\end{align}

\noindent Now we show that this equation can be satisfied for $a_1 >
0$.  The factor in square bracket on the right hand side is negative
for all $p^{\prime}$, so the right hand side of this equation is
positive.  And all factors on the left hand side of the equation are
positive, so the arbitrary constants $a_i$ can be positive or be in
such a combination to make at least the first term positive. This
shows that it is possible to have a dispersion relation in this case,
that has a phonon like term. The contribution of these modes of
excitation is present but is small in general. For sufficiently small
values of momentum however, the phonon like term would become
important.

This analysis can be extended for $D>6$ in the same way. The
dispersion relation in that case would be given by: $\epsilon (p) =
\epsilon_0 (p) + (1/p_D)[a_1p+ a_2p^2 + ...]$, with $\epsilon_0 (p)$
given as in Eq. (\ref{GenDisp}) and with all other terms suppressed
for a large value of momentum.
   
\acknowledgments

We thank Dawood Kothawala and Sudipta Sarkar for useful discussions
about the entropy of the horizon fluid in Lovelock theory Black
Holes. This work was supported by the Max Planck India Partner Group on
Gravity and Cosmology. J. L.  Lopez was supported by
CONACYT-M\'exico. S. S. is partially supported by Ramanujan Fellowship of
DST, India.


\begin{thebibliography}{cc}

\bibitem{BCH} J. Bardeen, B. Carter and S. Hawking, the four laws of black hole mechanics, Commun.\ Math.\
  Phys.\ {\bf 31}, 161 (1973)

\bibitem{Bek} J. Bekenstein, Black holes and Entropy, Phys.\
  Rev.\ D {\bf 7}, 2333 (1973).

\bibitem{Hawking1} S. W. Hawking, Particle creation by black holes, Commun.\ Math.\ Phys.\ {\bf 43}, 199
  (1975); Black holes and thermodynamics, Phys.\ Rev.\ D {\bf 13}, 191 (1976).

\bibitem{Damour} Thibault Damour, \emph{Surface Effects in Black Hole
    Physics, Proceedings of the second Grossman Meeting on General
    Relativity}, Ed. R. Ruffini p. 587, North Holland, (1982).

\bibitem{Chirco} G. Chirco, C. Eling and S. Liberati, Higher
    curvature gravity and the holographic fluid dual to flat
    spacetime, JHEP {\bf 1108} 009 (2011).

\bibitem{Bred1} I. Bredberg, A. Strominger \emph{et. al}, From
    Navier-Stokes to Einstein, JHEP.\ {\bf 1207} 146 (2012).

\bibitem{Bred2} I. Bredberg and A. Strominger, Black holes as
    incompressible fluids on the sphere, JHEP {\bf 1205}, 043 (2012).

\bibitem{Thorne} Kip S. Thorne \emph{et. al}, Black holes: The
    membrane paradigm, Yale University Press (1986).

\bibitem{Paddy} Dawood Kothawala and T. Padmanabhan, Entropy
    density of spacetime as a relic from quantum gravity, Phys.\
  Rev.\ D {\bf 90}, 12 (2014) 124060 . [arXiv:1405.4967]

\bibitem{Shanki1} Josef Skakala and S. Shankaranarayanan, Black
    hole thermodynamics as seen through a microscopic model of a
    relativistic Bose gas, Int.\ J. Mod.\ Phys.\ D {\bf 25}, 1650047, (2016).

\bibitem{Shanki2} Swastik Bhattacharya and S. Shankaranarayanan,
  Bekenstein-Hawking Entropy from criticality,
  [arXiv:hep-th/1411.7830], (2014).

\bibitem{Shanki3} Swastik Bhattacharya and S. Shankaranarayanan,
 How Emergent is Gravity?, Int. J. Mod. Phys. D, {\bf 24}, 1544005,
  (2015).

\bibitem{Dvali} G. Dvali and C. Gomez,  Black holes as critical
    point of quantum phase transition, Eur.\ Phys.\ J. {\bf 74} 2752
  (2014).

\bibitem{Lanc} C. Lanczos, Z.\ Phys.\ {\bf 73}, 147 (1932).

\bibitem{Lov} D. Lovelock, The Einstein tensors and its generalizations, J.\ Math.\ Phys.\ {\bf 12}, 498 (1971).

\bibitem{Zbach} B. Zwiebach, Curvature squared terms and string theories, Phys.\ Let.\ {\bf 156 B}, 315 (1985).

\bibitem{BoulDes} B. G. Boulware and S. Deser, String-Generated Gravity models, Phys.\ Rev.\ Lett.\
  {\bf 55}, 2656 (1985).
  
  \bibitem{Myers} Robert. C. Myers and Jonathan Z. Simon,
  Black-hole thermodynamics in Lovelock gravity, Phys.\ Rev. D
  {\bf 38}, 8 (1988)

\bibitem{Paddy1} Aseem Paranjape, Sudipta Sarkar and T. Padmanabhan,
 Thermodynamic route to Field equations in Lanczos-Lovelock
    Gravity, Phys.\ Rev.\ D {\bf 74}, 104015 (2006).

\bibitem{Paddy2} Sanved Kolekar, Dawood Kothawala and T. Padmanabhan,
 Two aspects of black hole entropy in Lanczos Lovelock models
    of gravity, Phys.\ Rev.\ D {\bf 85}, 064031 (2012).

\bibitem{Jacob} T. Jacobson, Arif Mohd and S. Sarkar, The
 Membrane paradigm of Gauss-Bonnet Gravity, [arXiv:1107.1260]

\bibitem{Kolekar} Sanved Kolekar and Dawood Kothawala, Membrane
 Paradigm and Horizon Thermodynamics in Lanczos-Lovelock gravity,
  JHEP {\bf{1202}}, 006 (2012).

\bibitem{Wilczek} Maulik K. Parikh and F. Wilczek, An action for
 black hole membranes, Phys.\ Rev.\ D {\bf 58}, 064011 (1998).
    
\bibitem{Wald} Robert M. Wald, Black hole entropy is the Noether
    charge, Phys.\ Rev.\ D {\bf 48}, 8 (1993).
    
\bibitem{Ron} Rong-Gen Cai, Gauss-Bonnet black holes in AdS
    spaces, \ Phys.\ Rev.\ D{\bf 65}, 084014 (2002),
  [arXiv:hep-th/0109133].

\bibitem{note1} We remind the reader that in the fluid description,
  the volume of the fluid is given by the area of the black hole
  horizon. That is the reason of using $A$ in our thermodynamic
  relations.
  
\bibitem{Thermo} Herbert B. Callen, Thermodynamics and an Introduction
  to Thermostatistics, John Wiley \& Sons (2006).

\bibitem{Katz3} Nathalie Deruelle, J. Katz and Sachiko Ogushi, Conserved charges in Einstein-Gauss-Bonnet theory, Class.\ Quant.\ Grav.\ {\bf 21} 1971 (2004).

\bibitem{Beig} R. Beig, Arnowitt-Deser-Misner energy and $g_{00}$, Phys.\ Lett.\ A {\bf 69} 153 (1978).

\bibitem{Ashtekar} A. Ashtekar and A. Magnon-Ashtekar, On conserved quantities in general relativity, 
J.\ Math.\ Phys.\ {\bf 20} 793 (1979).  

\bibitem{Laszlo} Laszlo B. Szabados, Quasi-local Energy-Momentum and angular momentum in GR: A review article, Living.\ Rev.\ {\bf 7} 4 (2004).

\bibitem{Katz1} J. Katz, A note on Komar's anomalous factor, Class.\ Quant.\ Grav.\ {\bf 2} 423 (1985).
%  Phys.\ {\bf 31}, 161 (1973).

\bibitem{Katz2} J. Katz, J. Bicak and D. Lynden-Bell, Relativistic conservation laws and integral constraints for large cosmological perturbations, Phys.\ Rev.\ D {\bf 55} 5957 (1997).

\bibitem{BCD1} J. D. Gunton and M. J. Bucckingham, {\it{Condensation
      of the ideal Bose Gas as a Cooperative Transition}}, phys.\
  Rev.\ {\bf{166}} 1, (1968).

\bibitem{BCD2} Ralf Beckham, Frithjof Karsch and David E. Miller,
  {\it{Bose-Einstein Condensation of a relativistic Bose Gas in d
      Dimensions}}, Phys.\ Rev.\ Lett.\ {\bf 43} 18, (1979).

\bibitem{Brustein}  R.~Brustein, A.~J.~M.~Medved, \ Phys.\ Rev.\ D {\bf 88}, 064010 (2013).      

\bibitem{Chamseddine} A. H. Chamseddine, Topological gauge theory of gravity in five-dimensions and all odd dimensions,
Phys. Lett. {\bf B233}, 291 (1989).

\bibitem{SFluids} Andreas Schmitt, "\emph{Introduction to
    superfluidity: Field theoretical approach and applications}",
  Lect.\ Notes \ Phys.\ {\bf 888} Springer (2015).
       
\bibitem{Landau} L. D. Landau and E. M. Lifshitz, \emph{Statistical
    Physics}, Part 1; Butterworth-Heinemann (1980).


\end{thebibliography}
\end{document}